\documentclass[journal=jacsat,manuscript=article]{achemso}
\usepackage[utf8]{inputenc}
\usepackage[T1]{fontenc}
\usepackage[version=3]{mhchem}
\usepackage{soul}
\usepackage{bm}
\usepackage{threeparttable}
\usepackage{color}


\makeatletter
\let\@fnsymbol\@arabic
\makeatother

\title{Extending the atomic decomposition and many-body representation, a chemistry-motivated monomer-centered approach for machine learning potentials}

\author{Qi Yu}
\affiliation{Department of Chemistry, Fudan University, Shanghai, 200438, P.R. China}
\email{qi_yu@fudan.edu.cn}

\author{Ruitao Ma}
\affiliation{Department of Chemistry, Fudan University, Shanghai, 200438, P.R. China}
\author{Chen Qu}
\affiliation{Independent Researcher, Toronto, Ontario M9B0E3, Canada}

\author{Riccardo Conte}
\affiliation{Dipartimento di Chimica, Universit\`{a} degli Studi di Milano, via Golgi 19, 20133 Milano, Italy}

\author{Apurba Nandi}
\affiliation{Department of Physics and Materials Science, University of Luxembourg, L-1511, Luxembourg City, Luxembourg.}

\author{Priyanka Pandey}
\affiliation{Department of Chemistry and Cherry L. Emerson Center for Scientific Computation, Emory University, Atlanta, Georgia 30322, U.S.A.}

\author{Paul L. Houston}
\affiliation{Department of Chemistry and Chemical Biology, Cornell University, Ithaca, New York
14853, U.S.A. and Department of Chemistry and Biochemistry, Georgia Institute of
Technology, Atlanta, Georgia 30332, U.S.A}

\author{Dong H. Zhang}
\email{zhangdh@dicp.ac.cn}
\affiliation{Department of Chemistry, Fudan University, Shanghai, 200438, P.R. China}
\alsoaffiliation{State Key Laboratory of Molecular Reaction Dynamics, Dalian Institute of Chemical Physics, Chinese Academy of Sciences, Dalian, 116023, P.R. China}

\author{Joel M. Bowman}
\email{jmbowma@emory.edu}
\affiliation{Department of Chemistry and Cherry L. Emerson Center for Scientific Computation, Emory University, Atlanta, Georgia 30322, U.S.A.}

\begin{document}

\clearpage
\begin{abstract}
Most widely used machine learned (ML) potentials for condensed phase applications rely on many-body permutationally invariant polynomial (PIP) or atom-centered neural networks (NN). However, these approaches often lack chemical interpretability in atomistic energy decomposition and the computational efficiency of traditional force fields has not been fully achieved. Here, we present a novel method that combines aspects of both approaches, and achieves state-of-the-art balance of accuracy and force field-level speed. This method utilizes a monomer-centered representation, where the potential energy is decomposed into the sum of chemically meaningful monomeric energies. Without sophisticated neural network design, the structural descriptors of monomers are described by 1-body and 2-body effective interactions, enforced by appropriate sets of PIPs as inputs to the feed forward NN. We demonstrate the performance of this method through systematic assessments of models for gas-phase water trimer, liquid water, and also liquid \ce{CO2}. The high accuracy, fast speed, and flexibility of this method provide a new route for constructing accurate ML potentials and enabling large-scale quantum and classical simulations for complex molecular systems.

%We demonstrate the precision and speed of this method for water using 1 and 2-b effective interactions among  water monomers trained on large data sets of CCSD(T) energies.  Permutational invariance is enforced by using appropriate sets of PIPs as the inputs to the NN. The NN is then trained on thousands of CCSD(T) energies and achieves excellent precision for applications to water and also a short demostration for \ce{CO2}.
\end{abstract}

%\flushbottom
%\thispagestyle{empty}
\clearpage
%\maketitle
%\newpage
\section{Introduction}
Computational simulations of molecular systems are essential for understanding complex processes in chemistry, biology, and material sciences. A significant challenge in both quantum and classical simulations is the extensive computation required for potential energy and force evaluations given molecular configurations. Direct \emph{ab initio} calculations using accurate electronic structure methods such as the ``gold standard'' coupled cluster theory with single, double, and perturbative triple excitations, CCSD(T), are ideal. However, it quickly becomes prohibitive for systems with more than 15 atoms. Although density functional theory (DFT) is widely used in \emph{ab initio} molecular dynamics simulations due to its relative efficiency, its limited accuracy and still unfavorable computational scaling present challenges for long-time simulations of large and complex systems.

Over the past two decades, machine learning potentials (MLP) have emerged as a promising approach to enable efficient and accurate computational simulations.\cite{gkeka2020machine,deringer2019machine,NN-2014,manzhos2020neural,meuwly2021machine,Braams2009,ARPC2018,PIP-NN-1,Guo16,FINN1,fu2023accurate,BPNN,behler2021four,chmiela2018towards,bartok2010gaussian,GP-2017-1,SchNet,PhysNet,DeepMD,EANN,NequIP,MACE,Allegro} For high-dimensional systems with tens of thousands atoms, such as condensed phase water, an atomistic representation of the potential is a popular choice:\cite{BPNN}
\begin{equation}
E_{\text{total}}=\sum_i^{N_{\text{atom}}}E_{i,\text{atomic}},
\end{equation}
where the total potential energy of the system $E_{\text{total}}$ is decomposed as the sum of atomic energies $E_{i,\text{atomic}}$ over all atoms. This Behler-Parrinello representation has been widely applied in various machine learning potentials. Typical examples include BPNN,\cite{BPNN} SchNet,\cite{SchNet} PhysNet,\cite{PhysNet} DeePMD\cite{DeepMD} and EANN\cite{EANN} etc.. Recent equivariant neural network potentials such as NequIP\cite{NequIP}, MACE\cite{MACE}, and Allegro\cite{Allegro} also employ the atomistic representation.

Analogous to the difference between atomic and molecular orbital energies in electronic structure theory, the concept of atomic energy in these ML potentials lacks effective chemical meaning, as the energy of the molecule, rather than individual atoms, is more relevant for understanding the molecular structural signatures and perturbations by its environment. Another aspect of the atomistic representation of potential energy is related to computational scaling, where the cost scales linearly with the total number of atoms in the system. It remains an open question whether this scaling can be further improved to achieve greater efficiency while maintaining the same or higher level of accuracy.

Another approach to obtain machine learning potentials for large molecular systems is the many-body representation, which has been widely used in the literature since the 1980s.\cite{gordon2012fragmentation,hodges1997contribution,dahlke2007electrostatically,WHBB,gora2011interaction,medders2015representation,yu2017communication,heindel2020many} Taking water potentials as examples, the most accurate ones, namely MB-pol,\cite{Mbpol23} q-AQUA,\cite{q-AQUA} and q-AQUA-pol,\cite{qAQUApol} use a many-body expansion for the total energy of $N$ water monomers:
\begin{equation}
  E_{\text{total}}=\sum_{i=1}^NE_{1-b}(i)+\sum_{i>j}^NE_{2-b}(i,j)+\sum_{i>j>k}^NE_{3-b}(i,j,k)+\sum_{i>j>k>l}^NE_{4-b}(i,j,k,l) + \cdots,
\end{equation}
where each term is obtained from training a machine learning potential (MLP) on the appropriate dataset. Specifically, the 1-body term represents the potential for the isolated water monomer, often modeled using the spectroscopically accurate \emph{ab initio} based Partridge-Schwenke potential.\cite{PS} The 2-body term is an MLP fit to dimer interaction electronic energies, the 3-body term is an MLP fit to trimer interaction energies, and the 4-body term is an MLP fit to tetramer interaction energies. This many-body formulation allows the use of permutationally invariant polynomial based methods, such as PIP,\cite{ARPC2018} PIPNN,\cite{PIP-NN-1} and FINN,\cite{FINN1} to accurately describe the $n$-body interactions involving $n$ molecules.

Despite recent successes in simulating water properties from the gas phase to the liquid phase using many-body MLPs,\cite{WHBB,Mbpol23,q-AQUA,qAQUApol,DHZhangwater1} it is well-known that many-body representation suffers from the rapidly increasing number of 3-body, 4-body, and higher-order terms. Consequently, long-time simulations of relevant molecular systems are often prohibitive.

In this work, we introduce a novel machine learning framework that combines the strengths of both methods while mitigates their weaknesses. This new monomeric framework leverages a new representation of the system's potential energy in terms of monomer energies instead of atomic energies and employs molecular energies only at the 1-body and 2-body levels. Permutational invariance is enforced by using PIPs as inputs to NNs, describing molecule's structure and environment. This critical aspect of our work echoes the use of PIPs\cite{PIP-NN-1,PIP-NN-3} and later efficient Fundamental Invariants\cite{FINN1,Fu18} as inputs to NNs in applications to gas phase molecules.  We term this new approach MB-PIPNet. We demonstrate that this method achieves high accuracy across a variety of molecular systems, ranging from gas phase clusters (e.g., water trimer) to condensed phase systems (e.g., liquid water and \ce{CO2}).

Our findings indicate that many-body interactions, such as 3-body interactions, can be accurately described using only 1-body and 2-body PIP bases in the neural network descriptor. This discovery significantly enhances the efficiency of our framework, enabling fast computational simulations of complex condensed phase systems at the cost of conventional force field. Our framework exhibits excellent performance in molecular dynamics (MD) simulations of liquid water and achieves significantly better computational scaling compared to other atomistic machine learning models. Furthermore, our new framework can be systematically extended to various types of molecular systems. Related challenges and possible solutions are also discussed.\\

\section{Results}
\subsection{Monomeric neural network model}
\begin{figure}[H]
\begin{center}
\includegraphics[width=1.0\textwidth]{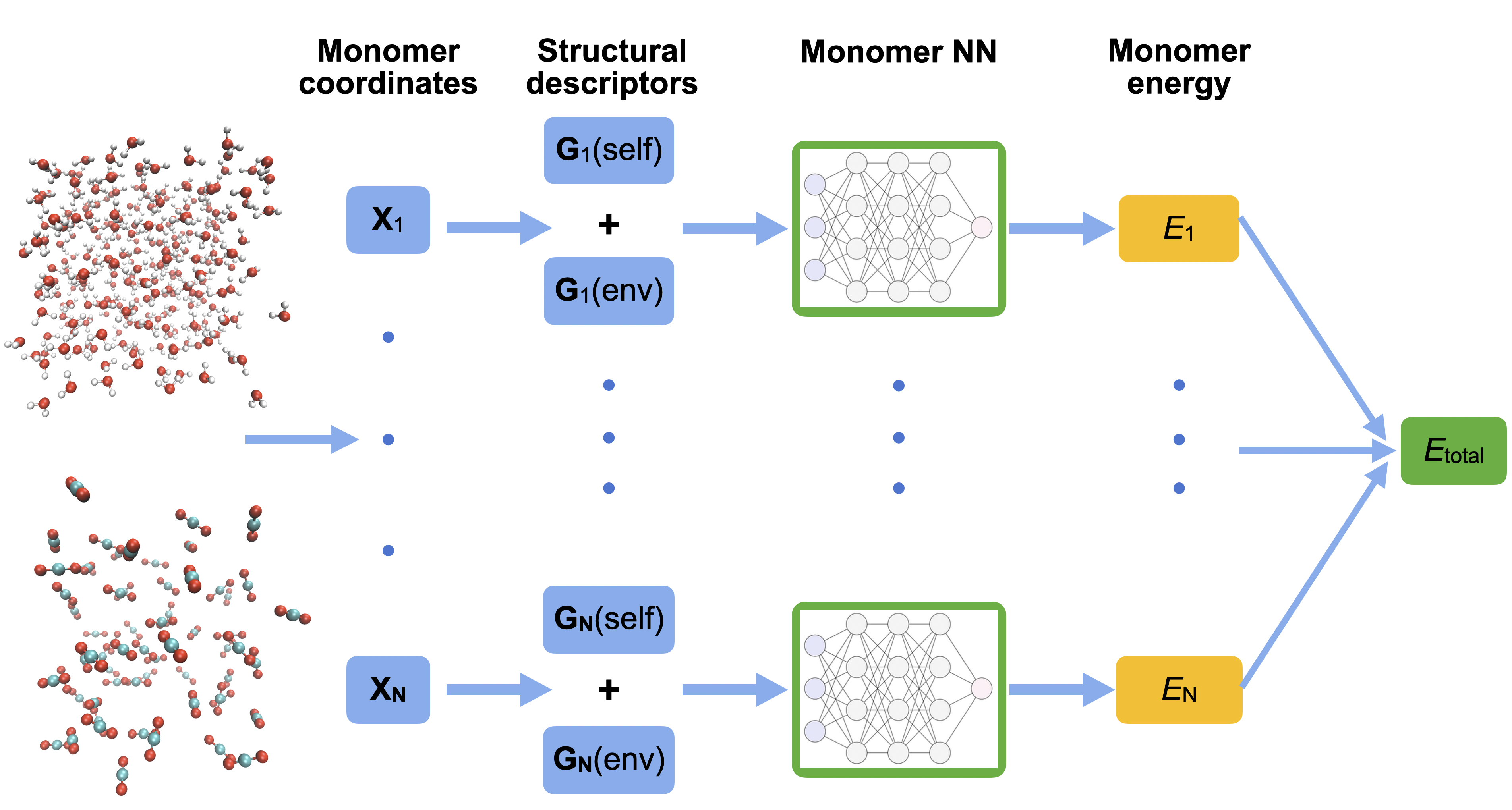}
\end{center}
\caption{\textbf{Schematic of the MB-PIPNet architecture.} The coordinates of each fragmental monomer is first transferred to this monomer's self-structural descriptors $\textbf{G}_i$(self), i.e., 1-body permutationally invariant polynomial (PIP) bases. The structural descriptors of each monomer's environment, $\textbf{G}_i$(env), are generated by pair-wise monomer coordinates involving different monomers, i.e., 2-body PIP bases. The self- and environmental-descriptors of each monomer are combined as the input of the neural network and yield the effective monomeric potential energy, $E_i$. The final energy of the complicated molecular system, $E_{\text{total}}$, is the sum over monomeric energies of all fragmental monomers.
}
\label{fig1}
\end{figure}

The proposed monomeric neural network potential model, MB-PIPNet, is illustrated in Fig. \ref{fig1}. This framework relies on appropriate decomposition of the molecular system into $N$ monomers. The total potential energy is then represented as the sum of perturbed energy of each monomer, $E_i$, analogous to the atom-centered approach, such that
\begin{equation}
E_{\text{total}}=\sum_i^{N}E_{i}
\end{equation}
The energy of each perturbed monomer is trained using a feed-forward neural network model that utilizes specifically designed structural descriptors as the input layer. In detail, after decomposing the entire molecular system into $N$ fragmental monomers, the Cartesian coordinates of monomer $i$ are transformed into a self-structural descriptor, $\textbf{G}_i(\text{self})$. We employ the widely used PIPs to construct this self-structural descriptor, which naturally ensures the necessary invariance to translation, rotation, and permutation. For instance, for a tetraatomic monomer, a set of symmetrized polynomials at order $n$ can be generated:
\begin{equation}
\textbf{G}_i(\text{self})=\textbf{P}(\textbf{X}_i)=\hat{\textbf{S}}[y_{12}^ay_{13}^by_{14}^cy_{23}^dy_{24}^ey_{34}^f]
\end{equation}
where $\hat{\textbf{S}}$ is the symmetrization operator that produces the appropriate sum of monomials with $a+b+c+d+e+f=n$. Each $y_{ij}$ is the Morse-like variable in terms of internuclear distance $r_{ij}$ between atom $i$ and $j$, such that $y_{ij}=\text{exp}(-r_{ij}/a_0)$ with $a_0$ as the hyperparameter. 

The self-structural descriptor, $\textbf{G}_i(\text{self})$, effectively describes the energetic response of a monomer to changes in its own configuration. However, each monomer is subjected to a complex environment with extensive intermolecular interactions involving other molecules. Consequently, the potential energy of each monomer should be polarizable. To account for this, we use 2-body PIPs as the environment descriptor for each monomer, $\textbf{G}_i(\text{env})$:
\begin{equation}
\textbf{G}_i(\text{env})=\sum_{j=1}^{N_{\text{mol}}\in R_{\text{c}}}\textbf{P}(\textbf{X}_i,\textbf{X}_j)f_c(\textbf{X}_i,\textbf{X}_j,R_\text{c})
\end{equation}
where $N_{\text{mol}}$ represents the total number of surrounding monomers within a distance cutoff of $R_\text{c}$. $\textbf{P}(\textbf{X}_i,\textbf{X}_j)$ is the corresponding 2-body PIPs generated from Cartesian coordinates of monomer $i$ and $j$. $f_c(\textbf{X}_i,\textbf{X}_j, R_\text{c})$ is a switching function that ensures a smooth transition to 0 for the 2-body polynomials when the distance between two monomers exceeds $R_\text{c}$. Taking water as example, the 2-body PIPs, $\textbf{P}(\textbf{X}_i,\textbf{X}_j)$, are generated with 42 symmetry for the \ce{H2O}$\cdots$\ce{H2O} pair. This includes permutational invariance for all four H atoms and both O atoms. These PIP bases are further purified to ensure they approach zero asymptotically as the distance between the oxygen atoms increases.\cite{q-AQUA}

The combination of $\textbf{G}_i(\text{self})$ and $\textbf{G}_i(\text{env})$ offers a systematic approach to describe the molecular response within a complex system. As will be explored below, using 1-b and 2-b PIPs as core components in these descriptors results in significantly more efficient computation compared to other ML methods. 
 
\newpage
\subsection{Energetic properties of MB-PIPNet model for water trimer}

\begin{figure}[H]
\begin{center}
\includegraphics[width=0.55\textwidth]{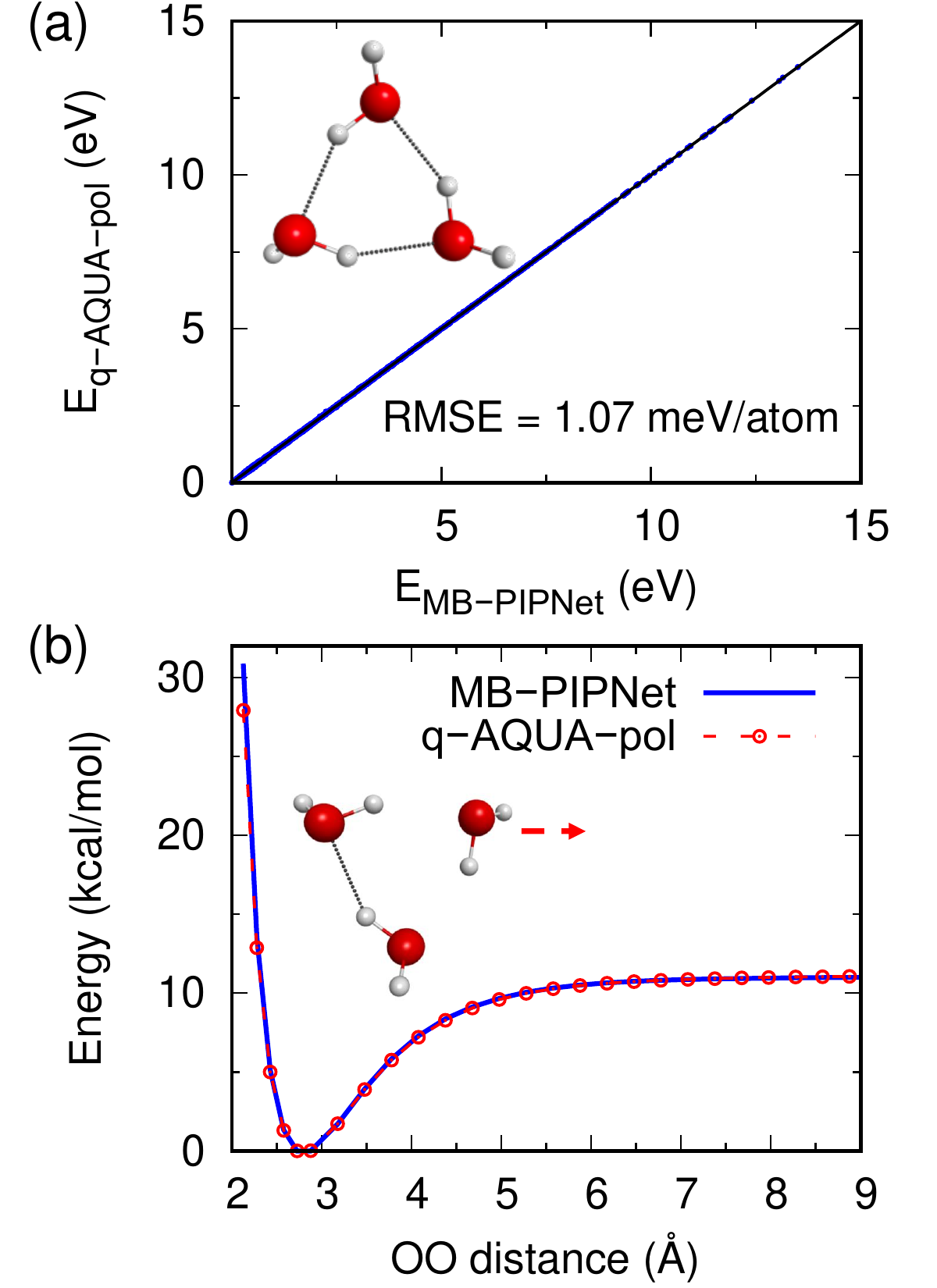}
\end{center}
\caption{\textbf{Potential energy predictions from MB-PIPNet model of water trimer.} (a) Energy-energy correlation plot for MB-PIPNet model of water trimer with reference energies calculated using q-AQUA-pol. (b) Potential energy curve predicted by MB-PIPNet model with comparison to q-AQUA-pol reference data.}
\label{fig2}
\end{figure}

The use of only 1-body and 2-body PIP bases as structural descriptors in our MB-PIPNet framework raises the question of whether MB-PIPNet can describe many-body interactions beyond two body. To address this, we first demonstrate MB-PIPNet's capability of capturing high-order interaction using the case of gas-phase water trimer. We trained an MB-PIPNet potential based on 45,812 trimer structures with energies calculated at the CCSD(T)-level q-AQUA-pol potential. This water trimer dataset spans a wide energy range of [0, 15] eV, and our final training root mean square error (RMSE) is only 1.08 meV/atom. As also shown in Fig. \ref{fig2}(a), the corresponding MB-PIPNet model for water trimer exhibits high accuracy in predicting potential energies of a separate test dataset, with RMSE of 1.07 meV/atom. We also examined the accuracy of the MB-PIPNet potential in a representative potential energy cut. As shown in Fig. \ref{fig2}(b), excellent agreement with the q-AQUA-pol reference data is achieved. These single-point energy results provide direct evidence that using only 1-body and 2-body PIP bases in constructing structural descriptors, the trained MB-PIPNet potential is capable of handling the complex molecular environments beyond simple 2-body interactions.

\begin{table}[H]
\footnotesize
\caption{Harmonic frequencies and anharmonic diffusion Monte Carlo (DMC) zero point energy (ZPE) (in cm$^{-1}$) of water trimer from different methods.}
\label{table1}

\begin{threeparttable}
    \begin{tabular*}{0.5\columnwidth}{@{\extracolsep{\fill}}l | c c c}
    \hline
    \hline
    \multicolumn{4}{c}{Harmonic frequency} \\
    \hline
    Method & MB-PIPNet & q-AQUA-pol & \emph{ab initio}\textsuperscript{a} \\
    \hline
    mode 1 &   155.4 & 160.4 & 154.5\\
    mode 2 &   174.7 & 176.5 & 178.6\\
    mode 3 &   188.9 & 188.3 & 185.7\\
    mode 4 &   195.7 & 194.3 & 191.7\\
    mode 5 &   219.7 & 220.1 & 220.2\\
    mode 6 &   229.1 & 237.5 & 228.3\\
    mode 7 &   329.2 & 337.0 & 332.4\\
    mode 8 &   351.9 & 350.8 & 346.4\\
    mode 9 &   438.4 & 437.3 & 437.1\\
    mode 10 &   555.4 & 562.1 & 558.8\\
    mode 11 &   648.9 & 653.8 & 656.8\\
    mode 12 &   829.6 & 844.6 & 846.5\\
    mode 13 &   1648.3 & 1662.2 & 1654.9\\
    mode 14 &   1655.3 & 1665.3 & 1660.2\\
    mode 15 &   1674.3 & 1684.0 & 1678.9\\
    mode 16 &   3618.9 & 3621.1 & 3621.0\\
    mode 17 &   3675.0 & 3681.2 & 3677.6\\
    mode 18 &   3684.7 & 3689.3 & 3685.5\\
    mode 19 &   3903.8 & 3907.2 & 3903.3\\
    mode 20 &   3908.2 & 3910.2 & 3908.3\\
    mode 21 &   3909.9 & 3914.3 & 3908.8\\
    \hline
    \multicolumn{4}{c}{Harmonic ZPE} \\
    \hline
    Method & MB-PIPNet & q-AQUA-pol & \emph{ab initio}\textsuperscript{a} \\
    \hline
     & 15997.7 & 16048.8 & 16017.7\\
    \hline 
    \multicolumn{4}{c}{Anharmonic DMC ZPE} \\
    \hline
   Method & MB-PIPNet & q-AQUA-pol\textsuperscript{b} & WHBB\textsuperscript{c} \\
    \hline
     &   15593 $\pm$ 4 & 15616 $\pm$ 2  & 15587 $\pm$ 2\\
    \hline
    \end{tabular*}
    \begin{tablenotes}
    \item[a] CCSD(T)-F12a/aug-cc-pVTZ
    \item[b] From Ref. \citenum{qAQUApol}
    \item[c] From Ref. \citenum{whbb_dmc_trimer}
   \end{tablenotes}
\end{threeparttable}
\end{table}

\newpage
The accuracy of the trained MB-PIPNet potential for the water trimer is further verified through harmonic normal mode analysis and anharmonic diffusion Monte Carlo (DMC) calculations. As seen in Table \ref{table1}, the MB-PIPNet potential provides accurate harmonic frequencies for the water trimer's global minimum structure, with deviations mostly smaller than 5 cm$^{-1}$ compared to both q-AQUA-pol and CCSD(T)-F12a/aug-cc-pVTZ benchmark results. As a more stringent test of the accuracy and smoothness of the PES, unconstrained DMC calculations were performed to determine the anharmonic zero-point energies (ZPE) of the water trimer using the trained MB-PIPNet potential. Notably, the rigorous ``exact'' quantum DMC calculations provide the exact ZPE of the molecule and also serve as an effective tool for detecting ``holes'' in the analytical PES. The trained MB-PIPNet potential was found to be ``hole''-free and the calculated water trimer's ZPE is $15593 \pm 4$ cm$^{-1}$ which agrees well with previous results using CCSD(T)-level PESs such as q-AQUA-pol and WHBB.

\subsection{Liquid water with MB-PIPNet potential}

Beyond gas-phase molecular clusters, it is crucial to assess the performance of the MB-PIPNet approach on condensed-phase systems, where each molecule is subject to a significantly more complex environment. To this end, we trained the MB-PIPNet model on a dataset consisting of 1,593 liquid water configurations, calculated at the revPBE0-D3 level of theory.\cite{cheng2019ab} Of this dataset, 90\% was randomly selected for training, with the remaining configurations used for testing. We employed the same 1-body and 2-body PIP bases as those used for the water trimer to generate the structural descriptors, $\textbf{G}$(self) and $\textbf{G}$(env). The RMSE errors on energy and force for the MB-PIPNet model, evaluated on the test set, are shown in Table \ref{table2}. In comparison to other MLPs, the MB-PIPNet model generally outperforms invariant atomistic MLPs, including BPNN and EANN. More sophisticated equivariant message-passing NN potentials, such as NequIP and MACE, exhibit better performance, particularly for force predictions. These MPNN models typically involve tens of thousands of parameters, suggesting that the RMSE error of the MB-PIPNet model could potentially be further reduced with more complicated NN structure and the incorporation of a message-passing mechanism.

\begin{table}[H]
\centering
\caption{Root mean square errors (RMSE) for energy  (meV/atom) and force ( meV/\AA) from different machine learning potentials trained on the same liquid water dataset from Ref. \citenum{cheng2019ab}.
}
\label{table2}
\begin{threeparttable}
\begin{tabular*}{1.0\columnwidth}{@{\extracolsep{\fill}} c c c c c c c }
\hline
\hline\noalign{\smallskip}
 & BPNN\cite{cheng2019ab} &  EANN\cite{PEANN} & NequIP\cite{NequIP} & MACE\cite{MACE,kovacs2023evaluation} & MB-PIPNet \\
\hline
Energy & 2.33  & 2.1 & 0.93 & 0.63 & 1.19 \\
Force & 120  & 129 & 45 & 36.2 &  93.3\\
\hline\noalign{\smallskip}
\hline
\end{tabular*}
\end{threeparttable}
\end{table}

\begin{figure}[H]
\begin{center}
\includegraphics[width=1.0\textwidth]{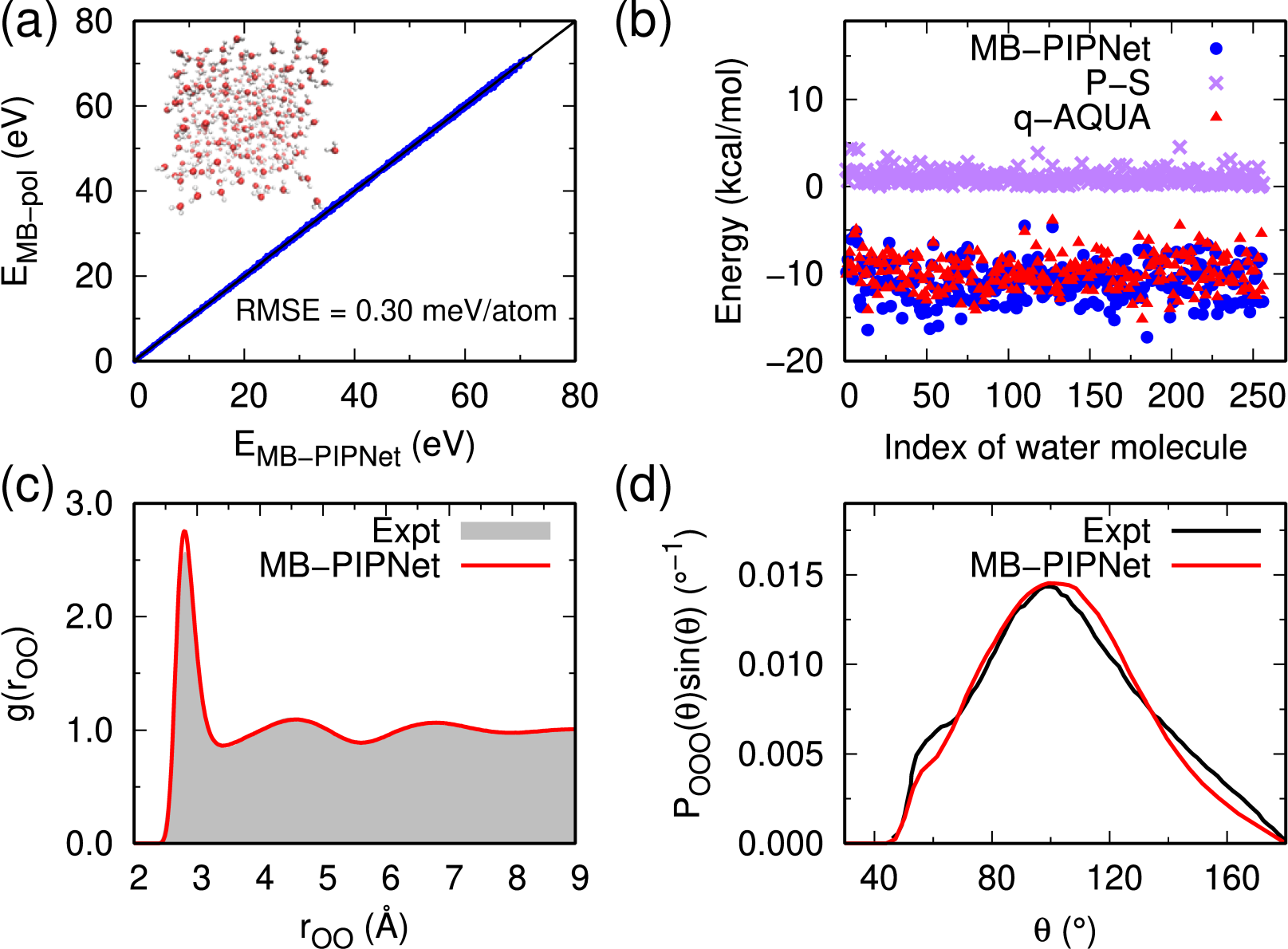}
\end{center}
\caption{\textbf{Potential energy predictions and molecular dynamics simulation results of liquid water by MB-PIPNet model.} (a) Energy-energy correlation plot for MB-PIPNet model of liquid water with reference energies calculated using MB-pol.
(b) Scatter plot of monomer energies of 256 water molecules in a periodic cubic box predicted by MB-PIPNet model, Partridge-Schwenke (P-S) water monomer potential, and q-AQUA model. (c) OO radial distribution function for liquid water at 298 K from classical MD simulations using MB-PIPNet model. The experimental data are taken from Ref. \citenum{Skinner2013,Skinner2014}. (d) The oxygen-oxygen-oxygen triplet angular distribution functions of liquid water at 298 K predicted by MB-PIPNet model. The experimental data are taken from Ref. \citenum{Soper2008}. The triplet angular distribution functions shown here were normalized to $\int_0^{\pi}\text{P}_{\text{OOO}}(\theta)\text{sin}(\theta)d\theta $.}
\label{fig3}
\end{figure}

To obtain a MLP model for liquid water with higher accuracy than density functional theory, we trained another MB-PIPNet model of water using using reference data from Zhai et al.\cite{zhai2023short} The training set consists of 75,874 different configurations from MD simulations of liquid water at various temperatures, using a cubic box of 256 water molecules under periodic boundary conditions. The total energy of each configuration was calculated using the MB-pol force field.\cite{mbpolliquid} The training process over this extensive dataset converged quickly, as shown in Supplementary Fig. 1. The final training RMSE is only 0.29 meV/atom, which is notably smaller than the 0.39-0.44 meV/atom achieved using the DeePMD approach with the same dataset.\cite{zhai2023short} Fig. \ref{fig3}(a) shows the performance of the MB-PIPNet model on energy predictions for a separate test dataset, showing good correlations with a small RMSE of 0.30 meV/atom. These energetic results verify the capability of the MB-PIPNet approach in handling complex and polarizable condensed phase systems, such as liquid water, where many-body interactions play crucial roles in determining the corresponding physical and chemical properties.

Conventional Behler-Parrinello-type atomistic MLPs predict the atomic energies of molecular systems. However, from a chemistry perspective, the energy of individual molecule is often of greater interest. A natural advantage of the MB-PIPNet model is its ability to directly predict the perturbed monomer energies of all individual molecules. This is analogous to the widely used concept of molecular orbital energy against atomic orbital energy in electronic structure theory. Fig. \ref{fig3}(b) presents the scatter plot of the monomer energies of 256 water molecules from a representative liquid water configuration using different methods. Given the coordinates of all water molecules, the Partridge-Schwenke (P-S)\cite{PS} energy of each molecule is calculated using corresponding spectroscopically accurate water monomer potential. The q-AQUA energies are calculated through many-body expansion using our recently developed purely many-body PES for water, where the energy of each water molecule is calculated as,
\begin{equation}
E_i(\text{q-AQUA})=E_{1-b}(i)+\sum_{j}^N\frac{1}{2}E_{2-b}(i,j)+\sum_{j>k}^N\frac{1}{3}E_{3-b}(i,j,k)+\sum_{j>k>l}^N\frac{1}{4}E_{4-b}(i,j,k,l)
\end{equation}
As seen, the P-S 1-body energies of 256 water molecules range from [0,5] kcal/mol, indicating the distorted structures of these molecules in the liquid phase relative to the global minimum structure.
When interactions among molecules are included, the water molecules are polarized and the corresponding q-AQUA monomer energies range from [-20,-3] kcal/mol. In line with the q-AQUA results, the MB-PIPNet model provides reasonable predictions of monomer energies, with different water molecules display distinct perturbed energies due to their structural distortion and interactions with other molecules in the liquid phase. These observations provide additional evidence that the MB-PIPNet model reasonably describes the many-body interactions in complex molecular systems with structural descriptors constructed from only 1-body and 2-body PIP bases. 

\begin{figure}[H]
\begin{center}
\includegraphics[width=0.5\textwidth]{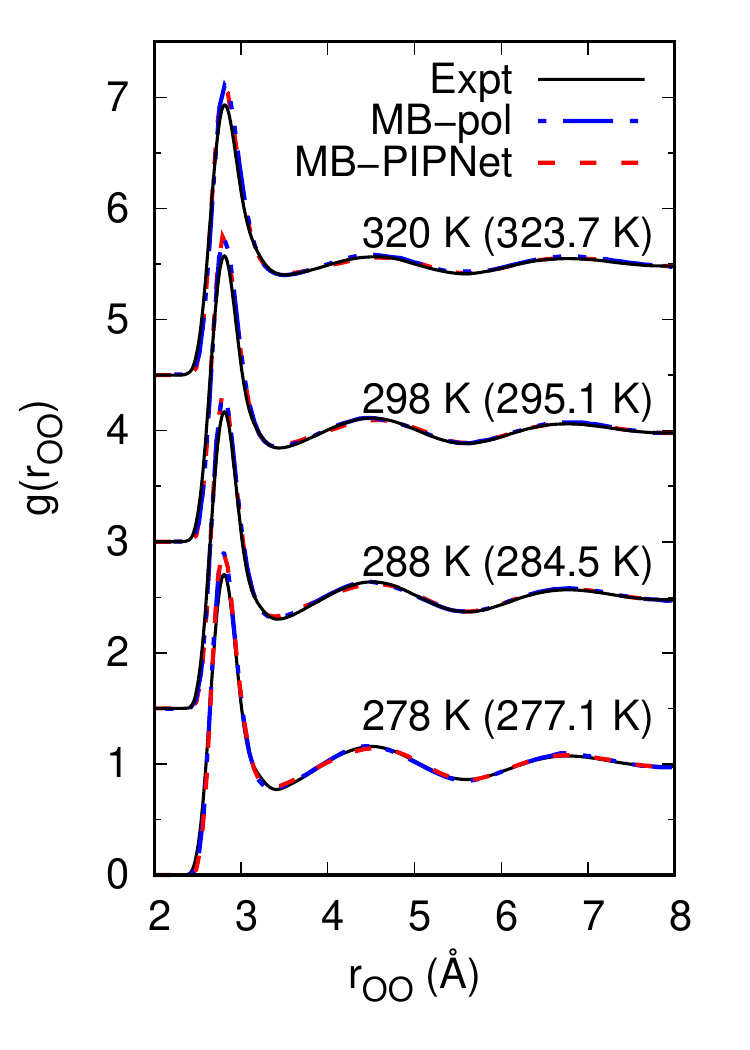}
\end{center}
\caption{\textbf{Structural properties of liquid water at different temperatures predicted by MB-PIPNet model.} OO radial distribution function from classical molecular dynamics simulations at different temperatures using MB-PIPNet model. The MB-pol data are taken from Ref. \citenum{mbpoltests}. The experimental data are taken from Ref. \citenum{Skinner2013,Skinner2014}}
\label{fig4}
\end{figure}

The trained MB-PIPNet model of liquid water was further employed in molecular dynamics simulations for bulk water properties using the i-PI software.\cite{i-pi} Fig. \ref{fig3}(c) shows the oxygen-oxygen (OO) radial distribution function (RDF) obtained from classical molecular dynamics simulations at 298 K, with the oxygen-hydrogen (OH) and hydrogen-hydrogen (HH) RDFs provided in Supplementary Fig. 2. As seen, the OO RDF obtained from the MB-PIPNet model agrees well with experimental data in terms of both peak positions and amplitudes. Fig. \ref{fig4} presents additional OO RDFs results generated by the MB-PIPNet and MB-pol across a range of temperatures. Both models consistently demonstrate excellent agreement with experimental data, highlighting the MB-PIPNet model’s capability to accurately replicate the MB-pol force field in simulating the structural properties of water across different thermal conditions.

%Excellent agreement between MB-PIPNet model-based OO RDFs and experimental data is also observed in MD simulations of liquid water over a range of temperatures, as shown in Fig. \ref{fig4}.
The oxygen-oxygen-oxygen triplet angular distribution P$_{\text{OOO}}$($\theta$) is another static property used to detect the tetrahedral orientational ordering of liquid water induced by the H-bonded network. We obtain P$_{\text{OOO}}$($\theta$) by computing the angle formed by an oxygen atom of a water molecule and two of its oxygen neighbors, with the neighbors defined using a cutoff of 3.27 \AA$\ $ to yield an average oxygen-oxygen coordination number of around 4.\cite{qAQUApol} As shown in Fig. \ref{fig3}(d), the distribution of P$_{\text{OOO}}$($\theta$) from the MB-PIPNet model is in excellent agreement with experiment in terms of peak position, width, and intensity. 
%These structural properties of liquid water predicted by the MB-PIPNet model agree with those predicted by the MB-pol force field used in our training data. 

Finally, the dynamic property of liquid water, specifically the self-diffusion coefficient $D$ as a function of temperature, is investigated using MB-PIPNet model-based MD simulations. The self-diffusion coefficient $D$ is calculated based on the slope of mean square displacements over time. As seen in Table \ref{table3}, the predicted $D$ from the MB-PIPNet model agrees well with the experimental measurements across different temperatures. Similar performance is observed in previous MD simulations performed directly using MB-pol\cite{mbpoltests}, although there are slight differences when compared to the MB-PIPNet results. It should be noted that the current self-diffusion coefficients were calculated using a simulation box of 256 water molecules. An increase in $D$ is anticipated for an ``infinitely'' large box, using the correction formula widely used in the literature and also our previous work.\cite{mbpoltests,qAQUApol}

\begin{table}[H]
\centering
\caption{Self-diffusion coefficient $D$ (\AA$^2$/ps) of liquid water at different temperatures
}
\label{table3}
\begin{threeparttable}
\begin{tabular*}{0.7\columnwidth}{@{\extracolsep{\fill}} c c c c}
\hline
\hline\noalign{\smallskip}
Temperature (K) & MB-PIPNet &  MB-pol\textsuperscript{a} &  Expt.\textsuperscript{b}   \\
\hline
278 & 0.136 $\pm$ 0.005 & 0.140 & 0.131 \\
288 & 0.177 $\pm$ 0.008 & 0.194 & 0.177 \\
298 & 0.251 $\pm$ 0.020 & 0.234 & 0.230\\
320 & 0.372 $\pm$ 0.014 & 0.344 & 0.360 \\
\hline\noalign{\smallskip}
\hline
\end{tabular*}
\begin{tablenotes}
\item[a] from Ref. \citenum{mbpoltests} 
\item[b] from Ref. \citenum{Mills1973} and \citenum{Holz2000}
\end{tablenotes}
\end{threeparttable}
\end{table}

Before illustrating the computational efficiency of the MB-PIPNet approach, we investigate the effect of cut-off distance, $R_\text{c}$, in constructing the environment descriptor $\textbf{G}(\text{env})$ in Eq. 5. In addition to the MB-PIPNet model trained using $R_\text{c}=9$ \AA, we also trained a MB-PIPNet model with $R_\text{c}=15$ \AA, termed as MB-PIPNet-long. During the model training phase, we did not observe a significant decrease in the training error for MB-PIPNet-long compared to the short-range model with $R_\text{c}=9$ \AA. Comparisons of the OO radial distribution functions between the two models are shown in Supplementary Fig. 4, where no significant differences are observed. This suggests the 2-body cut-off distance of 9 \AA$\ $ is sufficient to incorporate most interactions in liquid water and long-range interactions are implicitly included during the training process. A more careful assessment of the long-range interaction in the MB-PIPNet model is required and is subjected to our future study.
 
\subsection{Computational scaling with force-field-level cost}
Thus far, we have demonstrated that the MB-PIPNet approach can accurately describe many-body interactions from gas-phase clusters to condensed phase systems. The chemistry-motivated architecture of the MB-PIPNet method naturally provides detailed monomeric energies rather than conventional atomistic energies. Another appealing feature of the MB-PIPNet method is its favorable scaling and computational cost. This stems from two main aspects. The first one is associated with the use of permutationally invariant polynomials as key components in structural descriptors. The generation of PIPs have been extensively verified to be systematic and efficient compared to other complicated ML descriptors.\cite{houston2024no} Second, our MB-PIPNet framework employs a novel representation of the total energy as the sum of monomer energies. Consequently, the computational cost of the MB-PIPNet potential scales linearly with the number of molecules rather than the number of atoms, as is the case with most MLPs.

\begin{figure}[H]
\begin{center}
\includegraphics[width=0.7\textwidth]{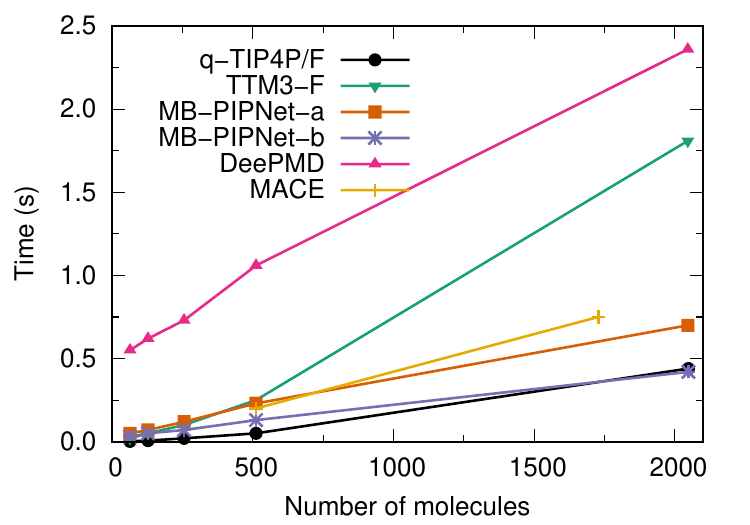}
\end{center}
\caption{\textbf{Computational cost of MB-PIPNet model.}
Computational time of single molecular dynamics step (energy and force) versus number of water molecules in a periodic simulation box using different methods. The timings for the MACE model\cite{kovacs2023evaluation} are obtained from Ref. \citenum{cheng2024cartesian} using an Nvidia A100 GPU. For other methods, all timing tests were performed using a single CPU core of the AMD EPYC 7002 processors. The MB-PIPNet-a and MB-PIPNet-b cases correspond to different 2-body cutoffs, $R_\text{c}=9$ \r{A} and 6 \r{A} respectively.}
\label{fig5}
\end{figure}

%need polishment for next paragraph
In Fig. \ref{fig5}, we compare the computational cost of the MB-PIPNet models with the q-TIP4P/F and TTM3-F force fields\cite{q-TIP4P, TTM3}, as well as the DeePMD and MACE MLPs\cite{zhai2023short}, for a single MD step of energy and gradient calculations across different sizes of liquid water simulation boxes. The MB-PIPNet-a and MB-PIPNet-b correspond to calculations with different 2-body cutoffs, $R_\text{c}=9$ \r{A} and 6 \r{A}, respectively. In the single CPU core simulations, we first verify the linear computational scaling of the MB-PIPNet approaches with respect to the number of water molecules. Moreover, with a larger 2-body cutoff, the MB-PIPNet-a model demonstrates computational efficiency comparable to the conventional polarizable water force field, TTM3-F, and is several times faster than the DeepMD potential. For simulations involving thousands of water molecules, the MB-PIPNet-a model significantly outperforms TTM3-F in speed, as it avoids the computationally expensive electrostatic Ewald summation. Similarly, the MB-PIPNet-b model, with a shorter $R_\text{c}=6$ \r{A}, shows even faster performance, surpassing TTM3-F and approaching the speed of the non-polarizable q-TIP4P/F force field. As the system size increases, the computational cost of MB-PIPNet-b becomes almost identical to that of q-TIP4P/F. Furthermore, when compared to the sophisticated equivariant message passing neural network, MACE, the MB-PIPNet-a model demonstrates similar or superior level of efficiency as the system size increases. Notably, this was achieved using only a single CPU core for the MB-PIPNet models, while the MACE timing tests were conducted on an Nvidia A100 GPU. Once extensive parallelization or GPU acceleration is applied, the MB-PIPNet approach is expected to achieve orders of magnitude improvements in computational efficiency than these MLPs.

As demonstrated above, our MB-PIPNet approach achieves a lower training RMSE than the invariant MLPs such as DeepMD using the same dataset while maintaining force field-level efficiency. Additionally, the MB-PIPNet potential is highly parallelizable, owing to its use of 1-body and 2-body PIPs for constructing structural descriptors and the formulation of monomeric energies. These features make it well-suited for microsecond-long simulations with first-principles accuracy of complex molecular systems. For instance, a recent study probing the liquid-liquid transition in supercooled water required long-term molecular dynamics (MD) simulations, spanning several years of GPU computational time, using the DeepMD-based potential \cite{sciortino2024pinpointing}. With our MB-PIPNet approach, it could be expected that at the same or higher level of accuracy, better statistical significant investigations of the unique properties of water could be conducted with significantly lower computational cost. Furthermore, in quantum mechanics/molecular mechanics (QM/MM) simulations of biomolecular systems, the MM region typically comprises tens of thousands of atoms—often 10,000 to 20,000 water molecules are required to properly solvate the biomolecule.\cite{senn2009qm} Instead of relying on conventional force fields, our MB-PIPNet approach opens new possibilities for biomolecular simulations at fully \emph{ab initio} levels of accuracy. While it is not yet feasible to perform \emph{ab initio} calculations for energy and forces on the entire system for training the model, as we will discuss later, the MB-PIPNet framework can be integrated with other strategies, such as many-body expansion, to construct accurate global PES for complex systems.

\subsection{Performance on other molecular systems and outlook}

\begin{figure}[H]
\begin{center}
\includegraphics[width=1.0\textwidth]{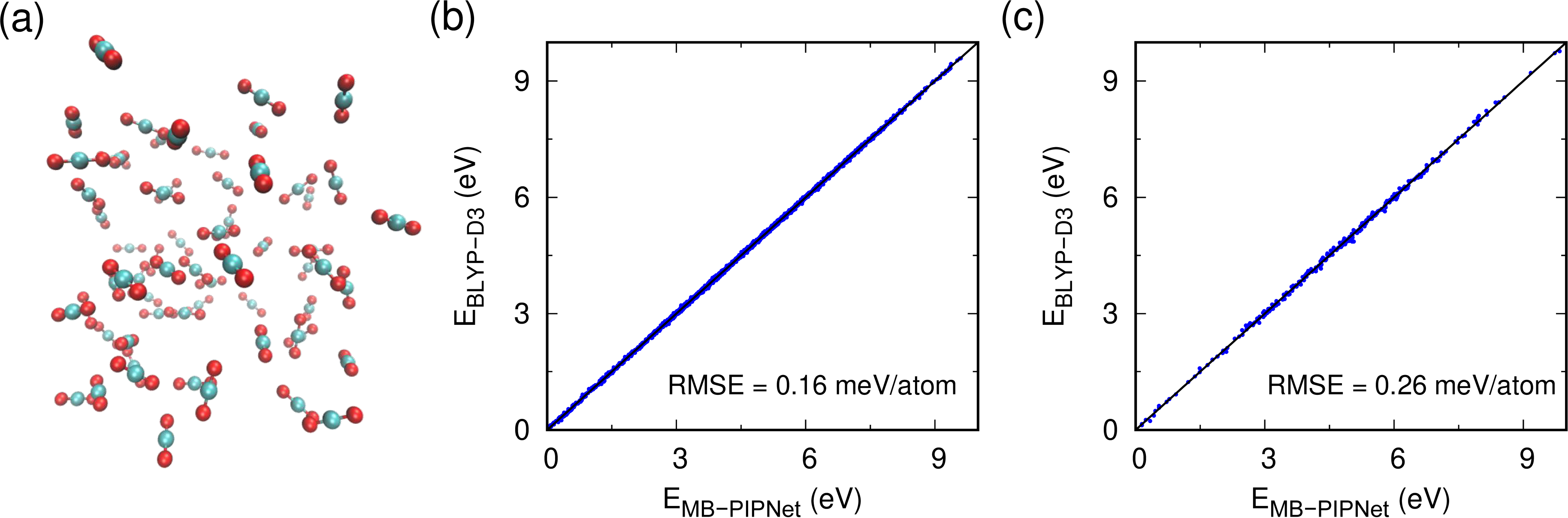}
\end{center}
\caption{\textbf{Performance of the MB-PIPNet model for liquid CO2.} (a) Schematic of liquid CO2 with 64 molecules in simulation box. (b)-(c) Correlation plots of training and test data sets with reference energies calculated at the BLYP-D3 level of theory.}
\label{fig6}
\end{figure}

The above two MB-PIPNet potentials are associated with gas-phase water trimer and liquid water. Here, we further demonstrate the transferability of the MB-PIPNet approach to different molecular systems. Fig. \ref{fig6} provides an example of MB-PIPNet model applied to liquid \ce{CO2}. The training data was obtained from Mathur et al.,\cite{mathur2023first} including configurations of 64 \ce{CO2} molecules in a simulation box. As detailed in the Methods section, the training process utilized only 2687 configurations calculated at the BLYP-D3 level of theory, and a small feed-forward NN structure was employed. Similar to the performance in other examples, the final training and test RMSE values are reasonably small, at 0.16 and 0.26 meV/atom, respectively. It should be noted that more thorough assessments of model training and MD-based property calculations are needed. This will be addressed in our future work by generating additional training data and conducting more systematic MB-PIPNet potential training. 

The MB-PIPNet framework is applicable to 
a wide variety of molecules and molecular interactions. This versatility stems from our PIP library, which has been instrumental in developing more than 60 MLPs for different molecules.\cite{houston2023pespip} Our PIP library can be directly interfaced with our MB-PIPNet approach to generate 1-body and 2-body terms as structural descriptors. However, the number of PIPs increases significantly with the number of atoms and polynomial orders, which often limits its application to systems with fewer than 15 atoms. Analogous to the behavior of FI-NN compared to PIP-NN, the fundamental invariant polynomials can also be employed as they are considered the minimal set of PIPs. The generation of FIs is systematic and has been frequently used in constructing high-dimensional MLPs. As noted in the recent work,\cite{chen2020fitting,fu2023accurate} an extension of FIs is anticipated for large systems with more than 20 atoms, or even 30 atoms. Extensive test employing FIs in our MB-PIPNet framework, in terms of accuracy and additional efficiency, is under investigation. 

\section{Discussion}
In this study, we present a novel framework for developing machine learning potentials for both gas-phase and condensed phase molecular systems. This framework, MB-PIPNet, discards the widely used atomic energy decomposition method and avoids the expensive computations of high-order terms in many-body representation. By representing the total potential energy as the sum of chemically meaningful monomeric energies, the MB-PIPNet framework accurately describes monomer structural distortion, environmental polarizations, and the final potential. It should be stressed that, unlike the standard many-body expansion approach, in MB-PIPNet, the combination of 1-body and 2-body PIP-based descriptors provides a consistent way to incorporate the perturbation effect of each monomer, where the perturbation results from the many-body interaction with other monomers. The MB-PIPNet framework greatly improves the computational cost of MLP evaluation, as the final computation scales linearly with the number of molecules instead of atoms. These features of the MB-PIPNet approach open new possibilities for performing computational simulations of complex systems such as molecular materials with first-principle accuracy but at the same speed as conventional force fields. To the best of our knowledge, such a balance between accuracy and efficiency has not yet been realized, even with atomic MLP like DeePMD.

The performance of MB-PIPNet approach has been systematically illustrated in quantum and classical simulations, i.e. quantum DMC and classical MD, of various molecular systems such as gas-phase water trimer and liquid water. As mentioned, generating molecular structural descriptors using PIPs or FIs is systematic and efficient, while ensuring rotational, translational, and permutational invariance. When training MB-PIPNet potentials for condensed phase systems like liquid water, a distance cutoff is incorporated for the 2-body PIPs of environmental descriptor $\textbf{G}(\text{env})$. Our tentative tests did not find significant differences between models with different cutoff thresholds. However, like other ML methods such as DeePMD, the current MB-PIPNet method lacks an explicit and robust description of the long-range effects. One possible direction is to incorporate a message passing mechanism for generating structural descriptors.\cite{gilmer2017neural,schutt2017quantum,EANN} 

Another advantage of the MB-PIPNet model is its flexible accuracy for the final PES. As shown in our examples and similar to other MLP approaches, the MB-PIPNet method can be directly used to train the total potential energy of the system. However the final accuracy of these MLPs heavily relies on the level of electronic structure theory, usually density functional theory, for generating the training data. This poses challenges in systematically elevating the accuracy of the MLP to methods like CCSD(T) as CC calculations are often prohibitive for larger systems. Conventional many-body representation of the potential is ideal for developing CC-level potential for condensed phase system like liquid water, but suffers from costly computation of high-order terms. Our MB-PIPNet framework offers new possibility for generating CC-level potential through a combination of many-body expansion. 
For example, in the case of liquid water, conventional 1-body and 2-body interactions can directly employ our recently developed high-accuracy CCSD(T)-level PESs in q-AQUA potential. For 3-body and higher-body interactions, the MB-PIPNet approach could 
directly fit the total $>2$-body energy contribution calculated from the q-AQUA PES.
Maintaining the accuracy of the final model, the combination of MB-PIPNet and many-body expansion is highly efficient without extensive calculations of higher-order terms and requires no additional structural descriptors for MB-PIPNet, as both methods reply on the same set of 1-b and 2-b PIP bases.

A final remark on our MB-PIPNet framework pertains to the need for reasonable assignments of fragmented monomers in complex molecular systems. It can be naturally applied to systems like gas-phase clusters and molecular liquids. However, additional efforts or further developments are needed for large organic molecules, biomolecules, or materials. Several possible directions include: (1) improving the relevant PIP or FI theory for efficient computation of polynomials in system with more than 20 atoms (2) developing fragmentation theory to automatically separate the large molecule into small fragments,\cite{qu2019fragmented} and (3) combining with atomistic ML methods for descriptions of material-molecule systems with broad applications in catalyst and material discovery. The monomer-centered concept of our MB-PIPNet can be further applied in other  machine learning approaches including invariant and equivariant neural networks.\cite{PhysNet,NequIP,MACE} We hope that the proposed new method will stimulate further development of MLPs in the wide fields of computational chemistry, physic, materials science, and biology for classical and quantum simulations of complex systems with \emph{ab initio}-level accuracy and conventional force field cost.

\section{Methods}
\subsection{Reference datasets}
For the water trimer, a total of 51,006 configurations were generated with energies calculated using the q-AQUA-pol potential. Among these, 45,332 configurations are trimer structures used in our previous three-body PES development in q-AQUA\cite{q-AQUA} and q-AQUA-pol\cite{qAQUApol} PESs. The remaining 5,674 structures were added by running diffusion Monte Carlo simulations using the initially trained MB-PIPNet models. The final dataset was randomly divided into a training dataset of 45,812 structures and a test data set of 5,194 structures. 

For liquid water, the first dataset is from Cheng et al.\cite{cheng2019ab} which includes 1593 liquid water configurations with each structure containing 64 water molecules. The energies were calculated at the revPBE0-D3 level of density functional theory. For the second dataset, we employed the one from Zhai et al.\cite{zhai2023short}. We refer readers to the original publication for more details. Briefly, a final training data set of 75,874 configurations and test dataset of 9,448 configurations were generated from MD simulations at different temperatures and pressure of 1 atm for a cubic box containing 256 molecules under periodic boundary conditions. The potential energy of each configuration was calculated from MB-pol force field.\cite{mbpol3b}  

The dataset for liquid \ce{CO2} was directly obtained from Mathur et al.\cite{mathur2023first} where the configurations were obtained from MD simulations of bulk liquid states at T = 220-300 K and P = 100 bar for a system of 64 \ce{CO2} molecules under periodic boundary condition. A total of 3,800 configurations at BLYP-D3 level of theory were used, with 2,687 for training and 313 for test.

\subsection{Training details}

For all MB-PIPNet models in this study, we employed 6th-order full-symmetry permutationally invariant polynomials for the self-structural descriptor, $\textbf{G}$(self), and 4-th order full-symmetry 2-body PIP bases for the environmental descriptor, $\textbf{G}$(env). Specifically, for both water trimer and liquid water, the input layer of NN has a dimension of 188, with 49 for $\textbf{G}$(self) and 139 for $\textbf{G}$(env). For liquid \ce{CO2}, the input layer is 80, with 30 for $\textbf{G}$(self) and 50 for $\textbf{G}$(env). All the MB-PIPNet models utilize NNs with two hidden layers. The neuron structures for these layers are [30,60] for the water trimer, [15,30] for liquid water, and [10,30] for liquid \ce{CO2} respectively. The training of all MB-PIPNet models was realized using the Levenberg-Marquardt algorithm.\cite{more2006levenberg} The training stopped when the learning rate dropped below 10$^{-5}$. The cutoff distances for the three systems were set as 9.0 \AA.
 
\subsection{Diffusion Monte Carlo simulations of water trimer}
We performed diffusion Monte Carlo (DMC) calculations which is considered as the standard approach to rigorously calculate the ground vibrational state wave function and the anharmonic ZPE in full dimensionality. This is based on the similarity between the diffusion equation and the imaginary-time Schr\"odinger  equation with an energy shift $E_\text{ref}$
\begin{equation}
    \frac{\partial \psi(\bm{x},\tau)}{\partial \tau} = 
    \sum_{i=1}^{N}\frac{\hbar^2}{2m_i}\nabla_i^2\psi(\bm{x},\tau)
    -\left[ V(\bm{x})-E_{\text{ref}}\right]\psi(\bm{x},\tau)
\end{equation}
The reference energy $E_\text{ref}$ in Eq. 7 is used to stabilize the diffusion system in its ground state, serving as an estimator for the zero-point energy.\cite{Anderson1975} We employed the unbiased, unconstrained implementation of DMC using 3$N$ Cartesians.\cite{Schulten} In this method, the DMC calculation begins with an initial guess of the ground-state wave function, represented by a population of $N(0)$ equally weighted Gaussian random walkers. These walkers then diffuse randomly in imaginary time according to a Gaussian distribution.  The population is controlled by a birth-death processes.\cite{Schulten} To maintain the number of random walkers around the initial value $N(0)$, $E_\text{ref}$ is adjusted at the end of each time step according to
\begin{equation}
    E_\text{ref}(\tau) = \langle V(\tau) \rangle - \alpha\frac{N(\tau)-N(0)}{N(0)}
\end{equation}
where $N(\tau)$ is the number of walkers at the time step $\tau$, $\alpha$ is a feedback parameter, typically around 0.1, and $\langle V(\tau) \rangle$ represents the average
potential energy of all of the walkers at that step. Finally the average of the $E_{ref}$ provides an estimate of the ZPE.

In this study, the DMC calculations were carried out for water trimer with the imaginary time step $\Delta\tau$ = 5 a.u. and $\alpha$ = 0.1. Five independent DMC calculations were performed. In each DMC calculation, the number of walkers is 40 000, and these walkers are equilibrated for 5000 time steps followed by 40 000 propagation steps. The statistical uncertainty is estimated as the standard deviation of the 5 DMC runs for the same system.

\subsection{Molecular dynamics simulations of liquid water}
We interfaced the MB-PIPNet water potential with the i-PI software\cite{i-pi} to enable classical molecular dynamics simulations to be performed for bulk water. The canonical ensemble (NVT) MD simulations were conducted with 256 water molecules in a periodically replicated simulation box with the experimental density set to be that at corresponding temperatures. For the classical MD simulation, at each temperature, we ran three independent 1 ns trajectories with time step of 0.25 fs. The static and dynamical properties were calculated as an average over the three trajectories. 

The self diffusion coefficient, $D$, of liquid water can be calculated from:
\begin{equation}
    D = \frac{1}{3}\int_0^{\infty} \langle \textbf{v}(0)\cdot \textbf{v}(t) \rangle dt = \frac{1}{6}\lim_{t \rightarrow \infty}\frac{d\langle \parallel \textbf{r}(t)-\textbf{r}(0)\parallel^2 \rangle}{dt}
\end{equation}
where $\langle \parallel \textbf{r}(t)-\textbf{r}(0)\parallel^2 \rangle$ is the mean square displacement (MSD). For each trajectory, we used the center of mass of each water molecule to calculate the MSDs and conducted linear fits to obtain the slope of the MSD curve. The self-diffusion constant $D$ is simply $1/6$ of the MSD slope and the final values reported in the main text are from the averaged values over different trajectories. 

\bibliography{ref}

\section{Author contributions}
Q.Y. conceived the project, performed calculations, and analyzed the data. R.M. performed timing tests. All authors contributed to writing the manuscript.

\section{Competing interests}
The authors declare no competing interests.

\section{Additional information}
Supplementary information is available.

\end{document}